\documentclass[10pt, conference, compsocconf]{IEEEtran}
\usepackage{epsfig,endnotes}
\usepackage[numbers,sort&compress]{natbib}
\usepackage{multirow}
\usepackage{subfig}
\usepackage{graphicx}

\newcounter{subcopyrightbox@save}
\usepackage{caption}
\usepackage{color, url}
\usepackage{xspace} 
\usepackage[ruled,linesnumbered]{algorithm2e}
\usepackage{mathrsfs}
\usepackage{amssymb}
\usepackage{amsmath}
\usepackage{epstopdf}
\usepackage{balance}

\newcommand{\myparatight}[1]{\smallskip\noindent{\bf {#1}:}~}
\newenvironment{packeditemize}{\begin{list}{$\bullet$}{\setlength{\itemsep}{0pt}\addtolength{\labelwidth}{10pt}\setlength{\leftmargin}{\labelwidth}\setlength{\listparindent}{\parindent}\setlength{\parsep}{0pt}\setlength{\topsep}{0pt}}}{\end{list}}

\renewcommand{\P}{A}
\newcommand{\A}{V}
\newcommand{\SADE}{ACTION}

\begin{document}

% Copyright
%\setcopyright{acmcopyright}
%\setcopyright{acmlicensed}
%\setcopyright{rightsretained}
%\setcopyright{usgov}
%\setcopyright{usgovmixed}
%\setcopyright{cagov}
%\setcopyright{cagovmixed}

%% DOI
%\doi{10.475/123_4}
%
%% ISBN
%\isbn{123-4567-24-567/08/06}
%
%%Conference
%\conferenceinfo{CCS'16}{October 24-28, 2016, Vienna, Austria}
%
%\acmPrice{\$15.00}
%
%
% --- Author Metadata here ---
%\conferenceinfo{WOODSTOCK}{'97 El Paso, Texas USA}
%\CopyrightYear{2007} % Allows default copyright year (20XX) to be over-ridden - IF NEED BE.
%\crdata{0-12345-67-8/90/01}  % Allows default copyright data (0-89791-88-6/97/05) to be over-ridden - IF NEED BE.
% --- End of Author Metadata ---
%\title{\bf \fontsize{15pt}{1em}\selectfont PIANO: Proximity-based Authentication for Internet-of-Things Devices}
\title{PIANO: Proximity-based User Authentication on Voice-Powered Internet-of-Things Devices\vspace{-10mm}}
%\numberofauthors{6} 

%\author{
%\alignauthor Neil Zhenqiang Gong\\
%       \affaddr{Iowa State University}\\
%       \email{neilgong@iastate.edu}
%\alignauthor Yu Wu \\
%       \affaddr{UC Davis}
%       \email{willwu0129@gmail.com}
%\alignauthor Richard Shin  \\
%       \affaddr{UC Berkeley}
%       \email{ricshin@berkeley.edu}
%\and
%\hspace{-20mm}
%\alignauthor \hspace{-6mm} Dawn Song  \\
%       \affaddr{\hspace{-6mm} UC Berkeley}
%      \email{\hspace{-6mm} dawnsong@cs.berkeley.edu}
%\alignauthor Hongxia Jin  \\
%       \affaddr{Samsung Research America}
%       \email{hongxia.jin@samsung.com}
%\alignauthor Xuan Bao   \\
%       \affaddr{Google Inc.}
%       \email{xbxuanbao8@gmail.com}
%}

\author{
    \IEEEauthorblockN{Neil Zhenqiang Gong\IEEEauthorrefmark{1}, Altay Ozen\IEEEauthorrefmark{1}, Yu Wu\IEEEauthorrefmark{2}, Xiaoyu Cao\IEEEauthorrefmark{1}, 
    Richard Shin\IEEEauthorrefmark{3}, Dawn Song\IEEEauthorrefmark{3}, Hongxia Jin\IEEEauthorrefmark{4}, Xuan Bao\IEEEauthorrefmark{5}}
    \IEEEauthorblockA{\IEEEauthorrefmark{1} Iowa State University,
%    \\\{1, 4\}@abc.com}
    {\IEEEauthorrefmark{2}UC Davis}, {\IEEEauthorrefmark{3}UC Berkeley}, 
         {\IEEEauthorrefmark{4} Samsung Research America},
         {\IEEEauthorrefmark{5} Google Inc.}}
}

%\IEEEoverridecommandlockouts
%\makeatletter\def\@IEEEpubidpullup{9\baselineskip}\makeatother
%\IEEEpubid{\parbox{\columnwidth}{Permission to freely reproduce all or part
%    of this paper for noncommercial purposes is granted provided that
%    copies bear this notice and the full citation on the first
%    page. Reproduction for commercial purposes is strictly prohibited
%    without the prior written consent of the Internet Society, the
%    first-named author (for reproduction of an entire paper only), and
%    the author's employer if the paper was prepared within the scope
%    of employment.  \\
%    NDSS '17, 21-24 February 2016, San Diego, CA, USA\\
%    Copyright 2016 Internet Society, ISBN 1-891562-41-X\\
%    http://dx.doi.org/10.14722/ndss.2016.23xxx
%}
%\hspace{\columnsep}\makebox[\columnwidth]{}}

\maketitle
\begin{abstract}

Voice is envisioned to be a popular way for humans to interact with Internet-of-Things (IoT) devices. 
%We call such IoT devices \emph{voice-powered} IoT devices.
We propose a proximity-based user authentication method (called PIANO) for access control on such \emph{voice-powered} IoT devices. %In this work, we focus on \emph{voice-powered} IoT devices and propose a novel proximity-based user authentication method (called PIANO) for voice-powered IoT devices. 
PIANO leverages the built-in speaker, microphone, and Bluetooth that voice-powered IoT devices often already have.  
Specifically, we assume that a user {carries} a personal voice-powered device (e.g., smartphone, smartwatch, or smartglass), which serves as the user's identity. When another voice-powered IoT device of the user requires authentication, PIANO estimates the distance between the two devices by playing and detecting certain {acoustic signals}; PIANO grants access if the estimated distance is no larger than a user-selected {threshold}. 
We implemented a proof-of-concept prototype of PIANO. Through theoretical and empirical evaluations, we find that PIANO is secure, reliable, personalizable, and efficient.

%PIANO uses a new acoustic signal based protocol (ACTION) that we design to estimate distances between the two devices. In PIANO, access to the authenticating device is granted if the estimated distance between the authenticating device and the vouching device is no larger than a user-selected \emph{authentication threshold}. We implemented a prototype of PIANO. Through extensive theoretical and empirical evaluations, we find that PIANO is secure, reliable, personalizable, passive, and efficient.  

%In PIANO, we assume that a user carries a personal IoT device, which serves as the user's identity. When another one of the user's IoT devices requires authentication, PIANO estimates the distance between the two devices by playing and detecting certain acoustic signals; PIANO grants access if the estimated distance is no larger than a user-selected authentication threshold. We implemented a prototype of PIANO. Through extensive theoretical and empirical evaluations, we find that PIANO is secure, reliable, passive, and efficient.  
\end{abstract}

\section{Introduction}
% Authentication is important.
% Existing method is not usable or insecure. 
% Wearable device is getting popular. Can we leverage wearable device for authentication.
% Our work proximity-based. previous work is inaccurate or insecure.  
% Our results: accurate, secure, and efficient. 

%Human authentication is one of the oldest security problems. 
%add reference on authentication on PC. 
%How to prevent unauthorized access on personal computers (PCs) and mobile devices is an important and fundamental research problem because users store a large amount of private information on them. For instance, on PCs, web browsers often store users' credentials to  web services such as emails and social networkings so that users can automatically log in these web services. Therefore, getting access to a user's PC allows an attacker to access the user's private information on all such web services.  Moreover, an attacker that obtains access to a smartphone can access the user's sensitive SMS messages, emails, and apps that the user didn't log out. 

%IoT is great. Access control on these devices is important. 
IoT devices are ever on the rise and getting ubiquitous. According to Gartner, the IoT devices installed base will grow to 26 billion and generate incremental revenue exceeding \$300 billion in 2020~\cite{gartner}.  
An attacker can compromise a user's security and privacy via
 \emph{unauthorized physical access} to the user's IoT devices.
%User authentication for IoT devices is very important. 
%Without a proper user authentication, 
%an attacker can have unauthorized physical access to a user's devices, resulting in severe security and privacy threats. 
Specifically, many IoT devices store various private information of the devices' owners. For instance, a user's smartphone or smartwatch might store the user's sensitive emails (e.g., emails including social security number and credit card numbers), sensitive messages, and call history. Likewise, a user's Bee (a healthcare monitoring IoT device~\cite{Bee}) stores the user's insulin injection data and glucose levels. Imagine a user leaves his/her such IoT devices unattended (e.g., when the user goes to restroom or to have lunch) or loses them, then an attacker could have unauthorized physical access to these devices to access the user's private information on them, compromising the user's data security. Moreover, having physical access to one IoT device could enable an attacker to control other connected IoT devices. For instance, smartphone is used to remotely control garage door~\cite{Smartgarage}; having access to a user's smartphone enables an attacker to remotely open the user's garage door via sending control commands to it, which could subsequently lead to house robbery.

Preventing unauthorized physical access often relies on \emph{user authentication}, i.e., we authenticate the identity of the user before allowing the access.
Existing user authentication methods such as \emph{password} and \emph{biometrics} are insufficient for IoT devices. 
Specifically, password and biometrics including fingerprint, face, and touch behaviors~\cite{frank2013touchalytics,GongTouchAsiaCCS16} are inapplicable to IoT devices that do not have a keyboard, touchscreen, fingerprint reader, or camera. Moreover, password is tedious, e.g., the legitimate user needs to input the password every time the user uses the device. Voice-based speaker recognition (also a biometric authentication) is applicable to voice-powered IoT devices, but voice-based speaker recognition, like other biometrics, is vulnerable to forgery attacks~\cite{jain201650}. For instance, an attacker can record the legitimate user's voice and replay it to bypass voice-based biometrics~\cite{aylett122008combining}.  Detecting forgery attacks is a challenging unsolved problem, though we have seen some progress in the past decade~\cite{jain201650,GongTouchAsiaCCS16}.

In this work, we focus on \emph{voice-powered} IoT devices and propose 
 a proximity-based user authentication method (PIANO) for voice-powered IoT devices. 
 Traditional human-computer interfaces such as keyboard and touchscreen have limited application in resource-constrained IoT devices, and voice is envisioned to be a popular way for humans to interact with IoT devices~\cite{voiceIoTExample,CarliniUsenixSecurity16}. Therefore, we focus on voice-powered IoT devices.
 Specifically, humans control a voice-powered IoT device via \emph{voice commands}, which are interpreted by the device via speech recognition techniques,
 and the device responds to humans via voice.  
 Such IoT devices are often equipped with built-in microphone and speaker in order to support voice-based interactions; and they are often also equipped with Bluetooth, a pervasive wireless communication technology, to exchange data with other IoT devices.  

Imagine a user Alice carries a smartwatch, which represents Alice's identity. When Alice uses her smartphone, the smartwatch and the smartphone are physically close; however, when an attacker tries to access Alice's smartphone while Alice is far away from her smartphone, the smartwatch and the smartphone are far away from each other. This scenario is not limited to smartwatch-smartphone pair. In fact, this is an emerging scenario in IoT: a user carries an IoT device which has already authenticated the user's identity. We call this device \emph{vouching device}. The user has another IoT device, which requires authentication before being used. We call this device \emph{authenticating device}. For instance, in our smartwatch-smartphone example, smartwatch is a vouching device and smartphone is an authenticating device. Likewise, the vouching device can be a user's smartphone and the authenticating device is the user's other voice-powered IoT device. 
We observe that when the legitimate user uses the authenticating device, the authenticating device and the vouching device are physically close. However, when an attacker tries to access the authenticating device while the legitimate user is away, the two devices are far away from each other. Therefore, in PIANO, access to the authenticating device is allowed if and only if the distance between the device and the vouching device is no larger than an \emph{authentication threshold}, e.g., 1 meter.

%In PIANO, we assume that the legitimate user \emph{carries} a voice-powered device, which 
% % that \emph{has already authenticated the user}. This device
%  represents the legitimate user's identity.  We call this device \emph{vouching device}. A vouching device could be the legitimate user's smartphone, smartwatch, smartglass, or a dedicated authentication token  that the legitimate user uses to prevent unauthorized physical access to his/her voice-powered IoT devices. 
%Suppose a voice-powered IoT device of the user requires user authentication. We call this IoT device \emph{authenticating device}.
%When the legitimate user uses the authenticating device, the authenticating device and the vouching device are physically close. However, when an attacker tries to use the authenticating device while the legitimate user is away, the authenticating device and the vouching device are physically far away from each other. Therefore, in our PIANO, access to the authenticating device is allowed if and only if the distance between the device and the vouching device is no larger than an \emph{authentication threshold}, e.g., 1 meter.  

A key component of  PIANO is to estimate distance between the vouching device and the authenticating device accurately, efficiently, and securely. 
Existing distance estimation protocols~\cite{distanceestimationSurvey,rasmussen2009proximity,RasmussenSecurity10,vasisht2016decimeter,sastry2003secure,peng2007beepbeep} are insecure or inaccurate.
% Estimating distance between two devices has attracted extensive studies in various communities. 
% % \cite{distanceestimationSurvey,kotaru2015spotfi,xiong2013arraytrack,vasisht2016decimeter,peng2007beepbeep,zhang2012swordfight,hightower2001location} and distance bounding~\cite{brands1993distance,hancke2005rfid,sastry2003secure,rasmussen2009proximity,RasmussenSecurity10,cremers2012distance} between two devices have been extensively studied in the security and pervasive computing communities. 
%However, existing distance estimation/bounding protocols~\cite{distanceestimationSurvey,kotaru2015spotfi,xiong2015tonetrack, brands1993distance,realizationDistanceBounding,vasisht2016decimeter,sastry2003secure, peng2007beepbeep,zhang2012swordfight} are insufficient for our proximity-based authentication because they are insecure or inaccurate. We discuss more details about existing distance estimation/bounding protocols in Section~\ref{related}. 
%they suffer from one or more of the following limitations: 1) they are insecure, 2) they are inaccurate,  3) they require specialized hardware, and 4) they only work in certain environments.  We discuss more details about existing distance estimation/bounding protocols in Section~\ref{related}. 
Therefore, 
we propose a new distance estimation protocol called \SADE. Our protocol leverages the speaker, microphone, and Bluetooth that voice-powered IoT devices often already have. 
%Specifically, voice-powered IoT devices are equipped with speaker/microphone to interact with humans via voice, and they are often equipped with Bluetooth to exchange data with other IoT devices. 
Specifically,  ACTION  pairs the authenticating device and the vouching device via Bluetooth, which involves human interaction but only needs to be done once. When calculating distance between the authenticating device and the vouching device, the authenticating device first generates two randomized acoustic \emph{reference signals} using a \emph{signal-construction algorithm} and transmits  them to the vouching device via a Bluetooth-based secure channel. Each device then uses a speaker to play a reference signal. Simultaneously, the two devices also record  signals using microphones. Next, each device computes the times when the two reference signals arrived at it using a \emph{signal-detection algorithm}. Finally, we estimate the distance by multiplying the \emph{speed of sound} and the time which sound takes to travel from one device to the other.

We implemented a prototype of PIANO on two smartphones. Via  theoretical and empirical evaluations, we find that PIANO has several promising features: \emph{secure}, \emph{reliable}, \emph{personalizable}, \emph{zero-interaction}, and \emph{efficient}. Specifically, PIANO has a very low probability of falsely accepting an attacker (i.e., secure), even if the attacker leverages various \emph{spoofing attacks} (we discuss them in Section~\ref{problem}) to manipulate our distance estimation protocol. PIANO achieves a low probability of falsely rejecting the legitimate user (i.e., reliable). Users can tune the authentication threshold to meet their personalized needs (i.e., personalizable). For instance, they can set the authentication threshold to be 0.5 meter if they are in an environment where 1 meter is too long to be safe.  
PIANO requires no actions from users in the process of authentication (i.e., zero-interaction). In our prototype, authentication can be finished within 3 seconds (i.e., computationally efficient). Moreover, using PIANO 100 times only consumes 0.6\% of the smartphone battery (i.e., energy efficient).

Our contributions can be summarized as follows:
%In summary, we make the following main contributions:
\begin{itemize}
\item We propose PIANO, a novel proximity-based user authentication method for voice-powered IoT devices.  

\item We design a new acoustic signal based distance estimation protocol (\SADE) to estimate distance between two devices accurately, efficiently, and securely.

%that is secure against spoofing attacks. 
%PIANO uses \SADE\ to estimate distances between two devices. 
\item We implemented a proof-of-concept prototype of PIANO. Via theoretical and empirical evaluations, we demonstrate that PIANO is secure, reliable, personalizable, and efficient. 
\end{itemize}

\section{Related Work}
\label{related}

\myparatight{Distance estimation protocols} The key component of PIANO is a protocol that estimates the distance between the vouching device and authenticating device.  Estimating distance between two devices has attracted much attention in various communities~\cite{distanceestimationSurvey,rasmussen2009proximity,RasmussenSecurity10,vasisht2016decimeter,sastry2003secure,peng2007beepbeep}. However, 
these protocols are \emph{insecure} or \emph{inaccurate}, making them insufficient for proximity-based authentication. We suspect the major reason why they are insecure is that they were not designed for security applications.

First, some protocols~\cite{distanceestimationSurvey,rasmussen2009proximity,RasmussenSecurity10,vasisht2016decimeter} leverage radio  signals such as  Wi-Fi, Bluetooth, and GSM.
%For instance, one recent study~\cite{} can accurately estimate distances between two devices via treating one device as a Wi-Fi access point and the other device as a Wi-Fi receiver.
Radio signals can go through a wall, which has serious implications for security when using them in proximity-based authentication. For instance, if the authenticating device and the legitimate user who carries the vouching device are in two different rooms that are separated by a wall, or they are in two different floors next to each other in the same building, then the two devices could still have a small distance. Therefore, in such scenarios, an attacker can easily access the authenticating device.

Second, some protocols~\cite{sastry2003secure,peng2007beepbeep} leverage acoustic signals.
%The key idea is that  the distance is the speed of sound multiples the time that sound takes to travel from one device to the other.
The key idea is that one device plays an acoustic signal; the other device detects when the acoustic signal arrives at it; and then the distance is the speed of sound multiples the time that the acoustic signal takes to travel from one device to the other.
Acoustic signals are less likely to go through a wall.
 However, they are vulnerable to spoofing attacks.
For instance, an attacker can simply replay the  acoustic signals to ``shorten" the distance for these protocols~\cite{sastry2003secure,peng2007beepbeep}. Moreover, some protocols (e.g., Echo~\cite{sastry2003secure}) require accurate estimation of a device's processing delay, which is challenging for IoT devices given IoT devices often have limited computing power. In other words, they are inaccurate on IoT devices. 
Our proposed protocol also leverages acoustic signals, but they will be secure against various spoofing attacks and they do not rely on accurate estimation of processing delays.

\myparatight{Determining proximity using ambient signals}
We note that some recent research~\cite{Varshavsky07,Shafagh14}  proposed to perform proximity-based authentication via checking whether the authenticating device and the vouching device are physically close based on their ambient signals. 
The  intuition is that  two  devices that are physically close   share similar  ambient radio signals, luminosity, and acoustic noise. 
These ambience-based approaches suffer from a few limitations. First, they aim to determine \emph{relative} distances but not \emph{absolute} distances between two devices. 
% Specifically, suppose we have three devices A, B, and C; these ambience-based approaches could determine which device (B or C) is closer to the device A based on their ambient signals. However, it is unclear how to calculate absolute distances between two devices from the ambient signals sensed by them. 
This limitation hurts the usability of the authentication system. 
For instance, some users might  want to set the authentication threshold to be 0.5 meter while some other users might want to set it to be 1 meter. 
It is unknown how these  ambience-based approaches can support such personalized user needs. On the contrary, our PIANO is \emph{personalizable}, allowing users to tune the authentication threshold for their personalized needs.
Second, these ambience-based approaches are insecure because attackers can modify the ambience around the two devices, e.g., attackers could play the same music around the two devices to modify their ambient acoustic signals. %For instance, suppose ambient acoustic signals are used to establish proximity; when the legitimate user, who carries the vouching device, is away from the authenticating device, attackers could play the same music around the legitimate user and the authenticating device, which enables the attackers to be authenticated into the authenticating device. 

\section{Problem Definition and Threat Model}
\label{problem}
\myparatight{Proximity-based authentication}
%We denote the vouching device and the authenticating device as $D_V$ and $D_A$, respectively. 
%Suppose a user carries a personal IoT device which has already authenticated the user's identity. 
%For instance, this IoT device could be a smartwatch which the user wears or a smartphone which has already authenticated the user. 
%We call this \emph{vouching device} and denote it as $D_V$. The user has another personal IoT device, which requires user authentication before being used. We call this \emph{authenticating device} and denote it as $D_A$.  For instance, $D_V$ could be a smartwatch which the user wears and $D_A$ could be the user's smartphone, or $D_V$ could be a smartphone which has already authenticated the user and $D_A$ could be a smart lock. 
%We observe that  the vouching device and the authenticating device  are physically close when the legitimate user uses the authenticating device, while they are far away from each other when an attacker tries to access the authenticating device. Therefore, 
%Suppose a user has a voice-powered IoT device that requires user authentication. We call this device \emph{authenticating device}. The user also carries another voice-powered IoT device (called \emph{vouching device}), which has authenticated the user and represents the user's identity.
In our proximity-based authentication, a user who tries to access the authenticating device is authenticated if and only if the distance between the authenticating device and the vouching device is no larger than a user-selected authentication threshold.
%We define the proximity-based authentication problem as follows. 
%\begin{definition}[Proximity-based Authentication] 
%The user who issues a voice command to the authenticating device is authenticated if and only if the distance between the authenticating device and the vouching device is no larger than an authentication threshold.
%\end{definition}
%
%The authentication threshold  can be set by users to meet their personalized needs. 
Intuitively, our proximity-based authentication propagates a user's identity from the vouching device to the authenticating device. 
%The key component of proximity-based authentication is to estimate the distance between two devices. 
%Our goal is to design a secure, reliable, and efficient proximity-based authentication system. To this end, we leverage acoustic signals since radio signals cannot estimate distances securely using commodity devices. 

\myparatight{Threat model}
%We consider that the user carries the vouching device. The vouching device and the authenticating device are not compromised by the attacker. 
We assume an attacker does not perform attacks to the authenticating device nor our authentication system when the legitimate user is physically close to the authenticating device. 
This is because an attacker faces the risks of exposing himself/herself in such scenarios. 
We consider the attacker's goal is to access the authenticating device when the legitimate user is away from the authenticating device (i.e., the distance between the authenticating device and the vouching device is larger than the authentication threshold). %For instance, suppose a user uses his/her smartphone as a vouching device; the user goes to have lunch and carries his/her smartphone, but the user leaves his/her voice-powered IoT device in a shared office; and an attacker tries to use the voice-powered IoT device while the user is away. 
Specifically, we consider the following two attacks. 
 \begin{packeditemize}
 %attacks
  %The success rates of such attacks are the FPRs of the authentication system. 
  \item {\bf Zero-effort attacks.} An attacker can directly try to use the authenticating device while the legitimate user is away. Due to  distance estimation errors, 
   the authenticating device would falsely authenticate the attacker with a certain probability. Since performing such attacks does not require much effort from the attacker,  we call them \emph{zero-effort attacks}. The success rates of zero-effort attacks are introduced by measurement errors from hardware and background noise. 

 \item {\bf Spoofing attacks.} In spoofing attacks, an attacker uses his/her own devices to play certain acoustic signals around the authenticating device and/or the vouching device, which aims to spoof the system to estimate the distance to be smaller than the authentication threshold. %For instance, one spoofing attack is to replay the acoustic signals used by the distance estimation protocol if they are fixed.  
 We will discuss specific spoofing attacks after we present our distance estimation protocol. 
  \end{packeditemize}

\section{Design of PIANO}
%\subsection{Overview}
\myparatight{Hardware requirements} PIANO requires the vouching device and authenticating device to be equipped with microphone, speaker, and Bluetooth. 
Voice-powered IoT devices often already have these hardware for functionality support. 
%A user can use his/her smartphone, smartwatch, or smart glass as a vouching device, which often has these hardware components. 
%Moreover, if our PIANO is widely adopted, we envision that industry might also manufacture dedicated small authentication tokens that have these hardware components. We note that built-in speaker, microphone, and Bluetooth are very cheap nowadays.  
%In this work, we focus on the scenarios where the authenticating device is a  voice-powered IoT device. Therefore, the authenticating device also already has these hardware components to support its functionality.  %In particular, microphone and speaker are needed to interact with humans via voice, and Bluetooth is a pervasive wireless communication technology used to exchange data with other IoT devices. 

\myparatight{Registration phase} In the registration phase, a user pairs the vouching device with the authenticating device using Bluetooth. This pairing process could involve human interactions, e.g., the user needs to manually confirm pairing between the two devices, but the pairing process only needs to be done once. After the two devices are paired, they can communicate securely via Bluetooth. 

\myparatight{Authentication phase} When a user tries to use the authenticating device, the authenticating device uses PIANO to verify the user's identity. %The  authenticating device executes the voice command if the user is authenticated. 
%PIANO requires the authenticating device and the vouching device to be equipped with a speaker and a microphone, and paired via a secure channel (e.g., Bluetooth). We note that  built-in speaker and microphone are very cheap nowadays (e.g., less than \$2 for one~\cite{speaker,microphone}).
%Figure~\ref{overview} shows an overview of the authentication phase of PIANO. When authentication on the authenticating device is required, 
Specifically, PIANO first checks whether the vouching device is still paired with the authenticating device via Bluetooth. If not, which often means that distance between the two devices is larger than the authentication threshold (we denote the authentication threshold as $\tau$), PIANO rejects the access; otherwise PIANO estimates the distance between the two devices using our distance estimation protocol called \SADE. %Our distance estimation protocol leverages the Bluetooth connection to exchange data securely. 
If the estimated distance is no larger than the authentication threshold, the access is granted, otherwise it is rejected. 

%Next, we will describe our distance estimation protocol. 
%We have shown that the basic distance estimation protocol is vulnerable to replay attacks, making it unsuitable for authentication. In the following, we enhance the basic protocol to be secure against the replay attacks, and we call the enhanced protocol \emph{secure distance estimation protocol (\SADE)}.  PIANO uses \SADE\ to estimate distances. 
%Moreover, we provide a formal security analysis about PIANO.

%In the following, we  introduce two acoustic signal based distance estimation protocols  with increasing complexity. We first describe a basic protocol, which is adapted from previous work. Second,  we show that this basic protocol is vulnerable to replay attacks (one type of spoofing attacks). Third,  we describe our secure protocol to defend against these attacks. Finally, we provide a formal security analysis about our secure protocol.

%\myparatight{Summary} We introduce two types of replay attacks to the basic protocol. In these  attacks, via replaying reference signals, an attacker  spoofs the authentication system to allow the attacker to access the authenticating device. These replay attacks succeed with a probability of $\frac{1}{4}$, which means that an attacker only needs to repeat the attacks for 4 times on average in order to get access to the authenticating device.  

\subsection{Overview of Our Distance Estimation Protocol}
%One solution to prevent the above replay attacks is to modify the cross-correlation algorithm to detect how many times a reference signal appears in the recorded signal. If a reference signal appears more than once, the authentication system requires a PIN or password. However, this modification allows an attacker to easily perform \emph{denial-of-service attacks}. Specifically, the attacker could repeatedly play the reference signals, which makes the authentication system plays 
%Figure~\ref{secureprotocol} shows the steps in our distance estimation protocol. 
Our protocol has the following steps.

\begin{packeditemize}
\item {\bf Step I:} The authenticating device constructs two snippets of acoustic signals, which we denote as $S_A$ and $S_V$, respectively. We call these acoustic signals \emph{reference signals}. 
\item {\bf Step II:} The authenticating device securely transmits the two reference signals $S_A$ and $S_V$ to the vouching device via Bluetooth. The communication channel is secure so an attacker cannot eavesdrop the reference signals. 

\item {\bf Step III:} The authenticating device and the vouching device record acoustic signals using a microphone. Moreover, the authenticating device uses a speaker to play the reference signal $S_A$ and the vouching device plays $S_V$.

\item {\bf Step IV:} The authenticating device detects when the two reference signals were recorded, and we denote the timestamps as $t_{AA}$ and $t_{AV}$, respectively. The vouching device also detects when the two reference signals were recorded, and we denote the timestamps as $t_{VA}$ and $t_{VV}$, respectively. %We note that the time when the reference signal $S_A$ was 

\item {\bf Step V:} The vouching device securely transmits the local time difference $(t_{VA}-t_{VV})$ to the authenticating device via Bluetooth.

\item {\bf Step VI:}  The authenticating device calculates the distance. 

%There are multiple ways to estimate the distance.
%First,  the distance is the speed of sound (denoted as $s$) multiplies the time that  the reference signal $S_A$ takes to travel from authenticating device to vouching device, i.e., $d_{A}=s \cdot (t_{VA} - t_{AA})$.
%Similarly, the distance can be estimated using the reference signal  $S_V$, i.e., $d_V=s\cdot (t_{AV} - t_{VV})$.
%However, these methods require the two devices to synchronize their time coordinates, and an error of 10 milliseconds in time synchronization could result in an error of more than 3 meters in distance estimation. To avoid time synchronization, we will estimate the distance $d_{AV}$ as  $d_{AV}=\frac{1}{2}(d_A+d_V) =\frac{1}{2}\cdot s \cdot ((t_{VA}-t_{VV}) + (t_{AV}-t_{AA}))$.
\end{packeditemize}

Next, we will elaborate Step I, Step IV, and Step VI.

% Compared to the basic protocol, we first add two new components, i.e., Step I and Step II. The authenticating device generates two randomized reference signals in Step I and sends them to the vouching device in Step II. Moreover, we design a new signal-detection algorithm to detect the reference signals  in Step IV. Note that \SADE\  requires the two devices to be already paired via a secure channel. This enables the references signals to be securely sent (a reference signal is an array of numbers, so they can be packed as data packets and sent over a secure wireless channel) to the vouching device in Step II, which means that attackers cannot eavesdrop nor modify the reference signals in Step II. Moreover, the difference between the arrival times of the two  reference signals that are played by the two devices can be securely sent to the authenticating device in Step V.

%\begin{figure}[t]
%\center
%%\vspace{-2mm}
%\includegraphics[width=0.45 \textwidth]{secureProtocol.pdf}
%%\vspace{-2mm}
%\caption{Illustration of our distance estimation protocol. }
%%Our new components are shown in bold. Compared to the basic protocol, we add new Step I and Step II, and we have a new signal-detection algorithm to estimate the arrival timestamps of reference signals in Step IV.}
%\label{secureprotocol}
%\vspace{-5mm}
%\end{figure}

\subsection{Step I: Constructing Reference Signals} 
\myparatight{Fixed vs. randomized reference signals} %We could construct fixed reference signals or randomized reference signals. Using fixed reference signals could be more efficient since they can be loaded to the two devices in the registration phase of PIANO, and then Step I and Step II of our distance estimation protocol can be eliminated. However, 
Using fixed reference signals makes the distance estimation protocol vulnerable to a very basic spoofing attack, i.e., \emph{replay attack}. Specifically, an attacker can obtain the fixed reference signals, e.g., via analyzing the implementation of PIANO. Then, in a replay attack, the attacker uses his/her own device to play the reference signals around the authenticating device or the vouching device such that the estimated distance is highly likely to be smaller than the authentication threshold. 
%Due to limited space, we omit the details of such replay attacks.
 We note that existing acoustic signal based distance estimation protocols~\cite{sastry2003secure,peng2007beepbeep} use fixed reference signals, and thus they are vulnerable to replay attacks. Therefore, we propose to randomize the reference signals every time authentication is required. 
 
 %We note that adding randomness is a popular method to prevent replay attacks in designing secure protocols, e.g., in the SSL/TLS session setup protocol, the client and the server exchange random numbers in order to establish a shared secret key. 

\myparatight{Frequency-domain randomized reference signals} It is challenging to randomize reference signals because how to randomize them also impacts the accuracy of detecting them. Specifically, we could construct randomized reference signals in either the time domain or the frequency domain.  For instance, one  way  is to construct an array of random numbers and treat it as a reference signal in the time domain. However, such randomized reference signals include a wide range of frequency components with random powers. As a result, these reference signals will be easily interfered by background noise, which makes detecting them inaccurate. 
%We should do some experimental results to show this.  
Therefore, we propose to construct randomized reference signals in the frequency domain. In particular, we discretize an appropriate frequency range (we will discuss the details of selecting the frequency range in Section~\ref{setup}) into $N$ bins and take the central point of each bin as candidate frequencies. We denote the $N$ candidate frequencies as a set $F_R$. 
To construct a reference signal, we first sample an integer $n$ ($0< n < N$) and then select $n$ frequencies from $F_R$ uniformly at random. For each sampled frequency, we synthesize a sine wave with the frequency, and then we construct a reference signal by adding these sine waves.

\subsection{Step IV: Detecting Reference Signals} 
In this step, both devices detect when the two reference signals arrived at them.  
Specifically, in Step III, each device has recorded a long audio sequence which includes the two reference signals. 
Then, in Step IV, each device detects the locations of the two reference signals in its recorded audio sequence, and translates the locations into timestamps.   
In the following, we take detecting one reference signal on the authenticating device as an example to illustrate our algorithm. Detecting the other reference signal is algorithmically the same, and the vouching device uses the same algorithm. 

In signal processing,  the \emph{cross-correlation algorithm} is a popular method to 
detect the location of a reference signal in a long signal sequence. 
For instance, BeepBeep~\cite{peng2007beepbeep} used this algorithm. 
%Specifically, the cross-correlation algorithm moves a window along the signal sequence with a certain step size. The length of the window equals the length of the reference signal.  For each window, the algorithm calculates the cross-correlation coefficient between the reference signal and the signal in the window. The index where the cross-correlation coefficient reaches its maximum is treated as the location of the reference signal in the signal sequence. 
Detecting our frequency-domain randomized reference signals using the cross-correlation algorithm results in high errors (see our experimental results in Section~\ref{comparewithCC}).  The key reason is a phenomena called \emph{frequency smoothing}, in which the power of a frequency component in a reference signal is distributed to nearby frequencies after the reference signal is played by one device and recorded by the other device.  %Figure~\ref{fs} shows the frequency smoothing phenomenon. %Due to the frequency smoothing effect, the powers of the frequencies in a frequency-domain randomized reference signal are distributed to nearby frequencies after the reference signal is played by a speaker and recorded by a microphone.
%The second reason is that the phase of a frequency component also changes after the reference signal is played and recorded. 
Due to frequency smoothing, after a reference signal $S$ is played by one device and recorded by the other device, its recorded version becomes $S'$, which is significantly different from $S$. However, the cross-correlation algorithm tries to detect $S$ in the recorded signal sequence, which results in high errors. %We note that it is hard to predict $S'$ from $S$ because their difference depends on the hardware on the devices. 

Therefore, we design a new algorithm to detect the locations of reference signals in the recorded signal sequence. Our algorithm leverages the frequency domain, and we call it \emph{frequency-based signal detection algorithm}.

\myparatight{Overview of our algorithm} Our core idea is to move a window along the recorded signal sequence; for each window, we obtain the power spectrum of the signal in the window; the window whose power spectrum best matches the power spectrum of the reference signal is treated as the location of the reference signal. 
Specifically, Algorithm~\ref{frequency} shows our frequency-based signal detection algorithm. Suppose we want to detect the location of a reference signal $S$ in a recorded signal sequence $X$. We denote the set of frequencies in $S$ as $F$, and we denote the power at each frequency $f\in F$ in the reference signal as $R_f$. 
 Our algorithm moves a window along the recorded signal with a step size $\delta$. For each window, we compute the \emph{normalized power} of frequencies in the reference signal. The index at which the normalized power reaches its maximum is treated as the location of the reference signal. When the reference signal is not in the recorded signal sequence, our algorithm outputs a special character $\bot$. Next, we discuss how we compute the normalized power.

\begin{algorithm}[t]
\DontPrintSemicolon
\KwIn{$X$, $S$, $F$, $R_f$ of frequency $f$ in $S$, and  $R=\{R_f|f\in F\}$. }
\KwOut{The location $l$ of $S$ in $X$.}
\Begin{
%//Coarse-grained search \;
$P_{max} = -\infty$ \;
\For{$i$ = 1 to $|X|-|S| + 1$ with a step size $\delta$}{
$P = NormPower(X[i \cdots i + |S| - 1], F, R)$ \; 
\If{$P > P_{max}$}{
	$P_{max}  = P$\;
	$l = i$\;
	}
}

//Reference signal $S$ is not in $X$ \;
$R_S = \sum_{f\in F} R_f$ \;
\If{$P_{max} < \epsilon R_S $}{
\label{lineepsilon}
	$l=\bot$\;
}
\Return $l$\;
			
%		//Fine-grained search \;
%		%$\alpha$ is used to determine the threshold.(0 $<$ $\alpha$ $<$ 1 ).
%		\For {$i=I-|S|$ to $I$ with a step size $s_f$} {
%			$A = NormalizedAmp(X[i \cdots i + |S| - 1], F)$ \; 		
%			\If{$A > \alpha \cdot A_{max}$}{
%				$L=i$\;
%				\Return $L$\;
%			}
%		}
}
\caption{Our Signal Detection Algorithm}
\label{frequency}
\end{algorithm}

\begin{algorithm}[!t]
\DontPrintSemicolon
\KwIn{$W$, $F$, and $R$. }
\KwOut{Normalized power of frequencies $F$ in $W$. }
		
\Begin{
$Y= PowerSpectrum(W)$\;

\For {$f\in F_R$}{
$i= \lfloor f/f_s\cdot |W| \rfloor$ \;
$P_f=\sum_{k=i-\theta}^{i+\theta} Y[k]$\;
\label{linef}
}

$P = -\infty$\;
//This sanity check enhances security \;
%//guessing-based replay attacks, defend against\;
%//all-frequency-based replay attacks, and\;
%//eliminate impact of background noise\;
\If{$P_f >\alpha R_f$ for all $f\in F$ and $P_{f'} < \beta$ for all $f'\in F_R \setminus F$}{
\label{condition}
	$P=\sum_{f\in F} P_f - \sum_{f'\in F_R \setminus F} P_{f'} $ \;
	\label{linenormpower}
}

\Return $P$\;

%\For {$f\in F$}{
%$i= \lfloor f/f_s\cdot |W| \rfloor$ \;
%$P_f=\sum\limits_{k=i-\theta}^{i+\theta} Y[k]$\;
%}
% - \frac{\theta}{\lambda}(\sum\limits_{k=f-\theta-\lambda}^{f-\theta} Y[k] + \sum\limits_{k=f+\theta}^{f+\theta +\lambda} Y[k])} $\;
		
%$P' = 0$\;
%\For {$f\in F_R \setminus F$}{
%$i= \lfloor f/f_s\cdot |W| \rfloor$ \;
%$P_f=\sum\limits_{k=i-\theta}^{i+\theta} Y[k]$\;
%
%$P' = P' + \sum\limits_{k=i-\theta}^{i+\theta} Y[k]  $\;
%}
%- \frac{\theta}{\lambda}(\sum\limits_{k=f-\theta-\lambda}^{f-\theta} Y[k] + \sum\limits_{k=f+\theta}^{f+\theta +\lambda} Y[k])} $\;

}
\caption{$NormPower$($W$, $F$, $R$)}
\label{power}
\end{algorithm}
%\vspace{-4mm}

\myparatight{Computing normalized power} Algorithm~\ref{power} shows how we compute the normalized power of frequencies $F$ of the reference signal in a given signal window $W$. 
In a nutshell, a signal window has a large normalized power if 1)  powers of the reference signal's frequencies in the window are comparable to those in the reference signal, and 2) the candidate frequencies that are not in the reference signal have small powers in the window. 
Specifically, we first get the power spectrum of the window via Fast Fourier transform (FFT). For each candidate frequency $f$ in the reference signal, we locate the index of $f$ in the power spectrum of the window (i.e., line 4); considering the frequency smoothing effect,  we compute the power of $f$ by aggregating the powers of the nearby $2\theta$ frequencies (i.e., line~\ref{linef}), where $\theta$ is the width of frequency smoothing. Then, if the power of each candidate frequency $f$ of the reference signal is larger than $\alpha R_f$ in the  window, and the power of each remaining candidate frequency that is not in the reference signal is smaller than a threshold $\beta$, we compute the normalized power of the window as the sum of the powers of frequencies in the reference signal minus that of the remaining candidate frequencies (i.e., line~\ref{linenormpower}), otherwise we treat the  normalized power of the window as a very small number, implying that the reference signal is not in the window.  Recall that $R_f$ is the power of frequency $f$ in the reference signal.

%If the reference signal appears in $W$, 1) the powers of the frequencies $F$ in $W$ should be comparable to those in the reference signal and 2) the powers of the remaining candidate frequencies $F_R \setminus F$  should be small enough. Therefore, if the power of each frequency $f$ in $F$ is larger than $\alpha R_f$ and the power of each remaining candidate frequency in $F_R \setminus F$ is smaller than a threshold $\beta$, we compute the normalized power as the sum of the powers of frequencies in $F$ minus that of the remaining frequencies in $F_R \setminus F$ (i.e., line~\ref{linenormpower}), otherwise we treat the  normalized power as a very small number, implying that the reference signal is not in the current window.  

Next, we explain why we introduce the parameters $\alpha$ and $\beta$. 
Reference signals are often \emph{attenuated} by hardware, i.e., after being played and recorded, a signal's powers become smaller. $\alpha$ is used to consider such attenuations. 
Suppose a background acoustic signal includes the frequencies in the reference signal and some other frequencies. If we do not perform  the sanity check about the powers of frequencies that are not in the reference signal (i.e., the sanity check using the threshold $\beta$ in line~\ref{condition}), then such a background noise could have a large normalized power, making detecting reference signals inaccurate. Likewise, an attacker can construct a spoofing reference signal via including all candidate frequencies and use it to perform replay attacks. If we do not perform sanity check in line~\ref{condition}, such spoofing reference signal will have a high normalized power, and our algorithm will  detect it as the reference signal,  making the corresponding replay attack  succeed with a high probability. 

%$\beta$ is a relatively small number and it is  used to eliminate the impact of background noise or an attacker's spoofing signal which might include the frequencies in a reference signal and some other candidate frequencies. For instance, an attacker could play a spoofing signal that includes all candidate frequencies; if we do not perform  the sanity check using the threshold $\beta$, the algorithm could detect the attacker's spoofing signal as the reference signal.

\myparatight{Reference signal is not present in the recorded signal sequence} In some scenarios, the authenticating device and the vouching device are far away from each other so that they cannot record each other's reference signals, e.g., the user who carries the vouching device goes to have lunch while leaving the authenticating device in a shared office.  Suppose an attacker tries to use the authenticating device  via performing zero-effort attacks. During the authentication process, the reference signal played by one device is not present in the signal recorded by the other device, and thus the maximum normalized power is not a reliable indicator of the reference signal, making distance estimation unreliable. To consider  such scenarios, our algorithm checks whether the maximum normalized power is smaller than  $\epsilon R_S$ (i.e., line \ref{lineepsilon} in Algorithm~\ref{frequency}), where $R_S$ is the power of the reference signal. If it is, our system outputs a special character and the authentication is denied.

\myparatight{Translating locations to timestamps} We denote by $l_{AA}$ and $l_{AV}$ respectively the detected locations of the reference signals $S_A$ and $S_V$ in the recorded signal sequence of the authenticating device. 
Moreover, we denote by $l_{VA}$ and $l_{VV}$ respectively the detected locations of the reference signals $S_A$ and $S_V$ in the recorded signal sequence of the vouching device. 
Suppose the location $l_{AA}$ corresponds to a time point $t_A$ on the authenticating device's time coordinate and the location $l_{VA}$ corresponds to a time point $t_V$ on the vouching device's time coordinate. Note that $t_A$ and $t_V$ are from two \emph{different} time coordinates. As we will show in the next section, our distance estimation does not rely on the specific values of $t_A$ and $t_V$. With these notations, we can transform the locations to timestamps as follows: $t_{AA} = t_A$, $t_{AV} = t_A + \frac{l_{AV}-l_{AA}}{f_A}$, $t_{VA} = t_V$, and $t_{VV} = t_V + \frac{l_{VV} - l_{VA}}{f_V}$, 
%\begin{align}
%\label{first}
%t_{AA} &= t_A \\
%t_{AV} &= t_A + \frac{l_{AV}-l_{AA}}{f_A} \\
%t_{VA} &= t_V \\
%\label{last}
%t_{VV} &= t_V + \frac{l_{VV} - l_{VA}}{f_V},
%\end{align}
where $f_A$ and $f_V$ are the sampling frequencies that the authenticating device's microphone and the vouching device's microphone use to acquire acoustic signals, respectively. 
Again, $t_{AA}$ and $t_{AV}$ are in the  time coordinate of the authenticating device; $t_{VA}$ and $t_{VV}$ are in the time  coordinate  of the vouching device; and the two time  coordinates could be different.

%simply denies the use without performing the rest steps of the protocol to estimate the distance.   
%\subsection{Step V: Compute Local Time Difference}

\subsection{Step VI: Estimating Distance}
In this step, the authenticating device estimates distance. There are multiple ways to estimate  distance between the two devices using the data we obtained in previous steps.  Specifically,
%\myparatight{Distance estimation using one reference signal}
 the distance between the two devices can be estimated by multiplying the speed of sound with the time that a reference signal takes to travel from one device to the other. In particular, given the timestamps $t_{AA}$ and $t_{VA}$,
%when $S_A$ arrived at the authenticating device, and the timestamp $t_{VA}$ when $S_A$ arrived at the vouching device,
 we can estimate the distance  as follows:
\begin{align}
\label{equ:dp}
d_A=s\cdot (t_{VA} - t_{AA}),
\end{align} 
where $s$ is the speed of sound. Alternatively, we can also estimate the distance using the  timestamps $t_{AV}$ and $t_{VV}$ when the reference signal $S_V$ arrived at the authenticating device and the vouching device, respectively. Formally, we have:
\begin{align}
\label{equ:da}
d_V=s\cdot (t_{AV} - t_{VV})
\end{align} 

However, using either Equation~\ref{equ:dp} or  Equation~\ref{equ:da} requires the two devices to synchronize their time coordinate systems.  For instance, Equation~\ref{equ:dp} requires that the timestamp $t_{VA}$ on the vouching device and the timestamp $t_{AA}$ on the authenticating device are measured from the same time coordinate system, which requires time synchronization on the two devices. However, time synchronization is generally a challenging task, and an error of 10 milliseconds in time synchronization could result in an error of more than 3 meters in distance estimation (speed of sound is around 340 $m/s$).%\footnote{\small This is because the speed of sound is around 340 $m/s$.} 

Therefore, we adopt a method developed in~\cite{peng2007beepbeep} to estimate distance, which avoids time synchronization. This method combines information about the two reference signals instead of a single one. Specifically, the method takes the average of the distances  in Equation~\ref{equ:dp} and  Equation~\ref{equ:da}. Formally, we have:
\begin{align}
\label{equ:d}
d_{AV}&=\frac{1}{2}(d_A+d_V) \nonumber \\ 
%&=\frac{1}{2}\cdot s \cdot ((t_{VA}-t_{VV}) + (t_{AV}-t_{AA})),
&=\frac{1}{2}\cdot s \cdot (- \frac{l_{VV} - l_{VA}}{f_V} + \frac{l_{AV}-l_{AA}}{f_A}).
\end{align} 
where $d_{AV}$ is the estimated distance between the two devices. 
%Via combining Equation~\ref{equ:d} with Equations~\ref{first} to Equations~\ref{last}, we have:
%\begin{align}
%\label{equ:dl}
%d_{AV}=\frac{1}{2}\cdot s \cdot (- \frac{l_{VV} - l_{VA}}{f_V} + \frac{l_{AV}-l_{AA}}{f_A}).
%\end{align} 
Equation~\ref{equ:d}  means that computing distance reduces to computing the location differences $(l_{VV} - l_{VA})$ and $(l_{AV}-l_{AA})$, which can be estimated by the two devices locally without time synchronization. %Therefore, in the protocol, the vouching device sends the time difference $(t_{AP}-t_{AA})$ to the authenticating device. 

\section{Security Against Spoofing Attacks}
\label{securityanalysis}
We assume that the attacker knows the candidate frequencies in $F_R$, and we consider the following two spoofing attacks.
\begin{packeditemize}
\item {\bf Guessing-based replay attacks.} An attacker could guess the reference signals and use them to perform replay attacks. Specifically, the attacker uses our signal construction algorithm to synthesize reference signals.  Performing a successful replay attack requires the attacker to guess the two reference signals correctly. 

\item {\bf All-frequency-based spoofing attacks.} An attacker can construct a spoofing reference signal that includes all candidate frequencies. Specifically, the attacker synthesizes a sine wave for each candidate frequency and constructs a spoofing reference signal by adding all these sine waves. Then, the attacker plays  the spoofing reference signal to  spoof our distance estimation protocol. 

\end{packeditemize}

\myparatight{Mitigating guessing-based replay attacks} We can defend against guessing-based replay attacks via using a relatively large set of candidate frequencies. 
If the frequencies in a spoofing reference signal do not match those in the legitimate reference signal, 
then the spoofing reference signal has a very small normalized power, and eventually our algorithm will output that the reference signal is not present in the recorded signal sequence, which means that the attacker is denied. 
%Recall that both types of replay attacks play the two reference signals at some times, and play random signals (e.g., a signal consisting of all candidate frequencies with powers larger than $\beta$) in the rest of time during the authentication process to jam the reference signals played by the authenticating device and the vouching device.  
The probability that  an attacker successfully guesses the  candidate frequencies in one reference signal is $\frac{1}{2^N-2}\approx \frac{1}{2^N}$. 
The replay attacks require guessing two reference signals correctly. Therefore, the probability that the attacker successfully performs a replay attack is $\frac{1}{2^{N+1}}$. 
When we use a relatively large number (e.g., 30) of candidate frequencies, the probability of successful attack is negligible. 
%For instance, when $N=30$, the probability to successfully perform a replay attack is $4.7\times 10^{-10}$, i.e., the attacker needs to repeat for $2.1 \times 10^{9}$ times on average to successfully perform a replay attack.

\myparatight{Mitigating all-frequency-based spoofing attacks} Suppose the attacker constructs a sine wave for each candidate frequency, and these sine waves have the same power $P_a$. Then, the attacker adds these sine waves to construct a long  signal and plays it in the entire authentication process. We can defend against such spoofing attacks via 
 constructing the reference signals with large enough powers such that $\alpha R_f$ is larger than $\beta$ (refer to the line \ref{condition} of Algorithm~\ref{power}). When $\alpha R_f>\beta$, for all windows except those that include the reference signal, the sanity check in the line \ref{condition} of Algorithm~\ref{power} fails no matter how the attacker chooses $P_a$. Specifically, if $P_a \geq \alpha R_f$, the sanity check about the powers of  the candidate frequencies that are not in the reference signal fails; if $P_a \leq \beta$, the sanity check about the powers of  the frequencies that are in the reference signal fails; and if $\beta < P_a < \alpha R_f$, both sanity checks fail.  As a result, our Algorithm~\ref{power} defines the normalized powers for these windows as negative infinity. Therefore, either the reference signal is still accurately detected or our algorithm reports that the reference signal is not present in the recorded signal. In either case, the attacker is denied access.

\section{Experiments}
\subsection{Experimental Setup}
\label{setup}
\myparatight{Proof-of-concept prototype} We implemented a prototype of PIANO as two Android apps and we run them on two Samsung Galaxy S4 smartphones; one is used as the authenticating device while the other is treated as the vouching device. The two smartphones are paired using Bluetooth. In our implementation of Algorithm~\ref{frequency}, we use adapted step sizes instead of using a fixed step size to achieve a trade-off between efficiency and accuracy. Specifically, we first use a step size of 1,000 to locate the window where the normalized power reaches the maximum; then we use a step size of 10 to perform a more fine-grained search around the window to locate a more accurate maximum.  Moreover, we detect the two reference signals simultaneously in one scan of the recorded signal, which is more efficient than detecting them in two scans. 

\myparatight{Reducing impact of background noise} Background acoustic signals could impact the accuracy of distance estimation since our protocol uses acoustic signals. Specifically, if a certain background noise has a power spectrum that is close to that of a reference signal, our distance estimation protocol might detect the background noise as the reference signal, which makes distance estimation inaccurate. 
We collected background acoustic noises in various environments (office, home, street, etc.) and found that most powers of background noises concentrate on frequencies that are smaller than around 6K Hz. Therefore, when we select candidate frequencies, we do not consider frequencies that are less than 6K Hz.

\myparatight{Using inaudible sound frequency} We set the sampling frequency on both smartphones to be 44.1K Hz, which is  the largest sampling frequency supported by the Android system. Given such sampling frequency and that background noises mainly concentrate on frequencies less than 6K Hz, the \emph{aliasing frequencies} of background noise mainly concentrate in the range  [38K Hz, 44K Hz]. Considering background noise, we use the frequency range  [25K Hz, 35K Hz]. Specifically, 
%Therefore, we consider frequencies in the frequency range $[6K, 20K]$.  
we equally divide this frequency range to be 30 bins and take the center of each bin as a candidate frequency, i.e., we have 30 candidate frequencies. We note that our constructed reference signals are almost inaudible (they are not completely inaudible due to hardware imperfection).

\myparatight{Setting parameters} 
When we construct a reference signal, we make its power the maximum value that can be represented by the Android system. Specifically, suppose we sampled $n$ candidate frequencies when constructing a reference signal. For each frequency $f$, we synthesize a sine wave with a power $(\frac{32000}{n})^2$ (i.e., $R_f=(\frac{32000}{n})^2$ in Algorithm~\ref{frequency} and Algorithm~\ref{power}), where we use 32000 because the Android system uses 16 bit integer to represent signals in the time domain. 
 Since FFT requires the length of the signal to be a power of 2, we set the length of our reference signals to be 4,096, which lasts for 93 milliseconds given our sampling frequency is 44.1K Hz.   
 We set $\epsilon=\alpha=1\%$ and $\beta=0.5\%\times R_f$ to be secure against various spoofing attacks. Moreover, we set $\theta=5$ to consider frequency smoothing effects.

% In FFT, we will get the second half of power spectrum to get the components whose frequencies are larger than the Nyquist frequency.

%In the following, we show the accuracy of distance estimation and accuracy of authentication decisions in various scenarios, and efficiency of PIANO. Since we have performed formal security analysis about spoofing attacks in Section~\ref{securityanalysis} and empirically evaluating them requires a large number of trials to trigger a successful spoofing attack, we focus on zero-effort attacks when we evaluate FPRs of PIANO.  

%\begin{figure}[t]
%\center
%%\vspace{-2mm}
%{\includegraphics[width=0.45 \textwidth]{realdistance.pdf}}
%%\vspace{-2mm}
%\caption{Illustration of measuring  real distances between two smartphones.}
%\label{realdistance}
%%\vspace{-4mm}
%\end{figure}

\begin{figure}[t]
\center
\vspace{-2mm}
\subfloat[In an office]{\includegraphics[width=0.25 \textwidth]{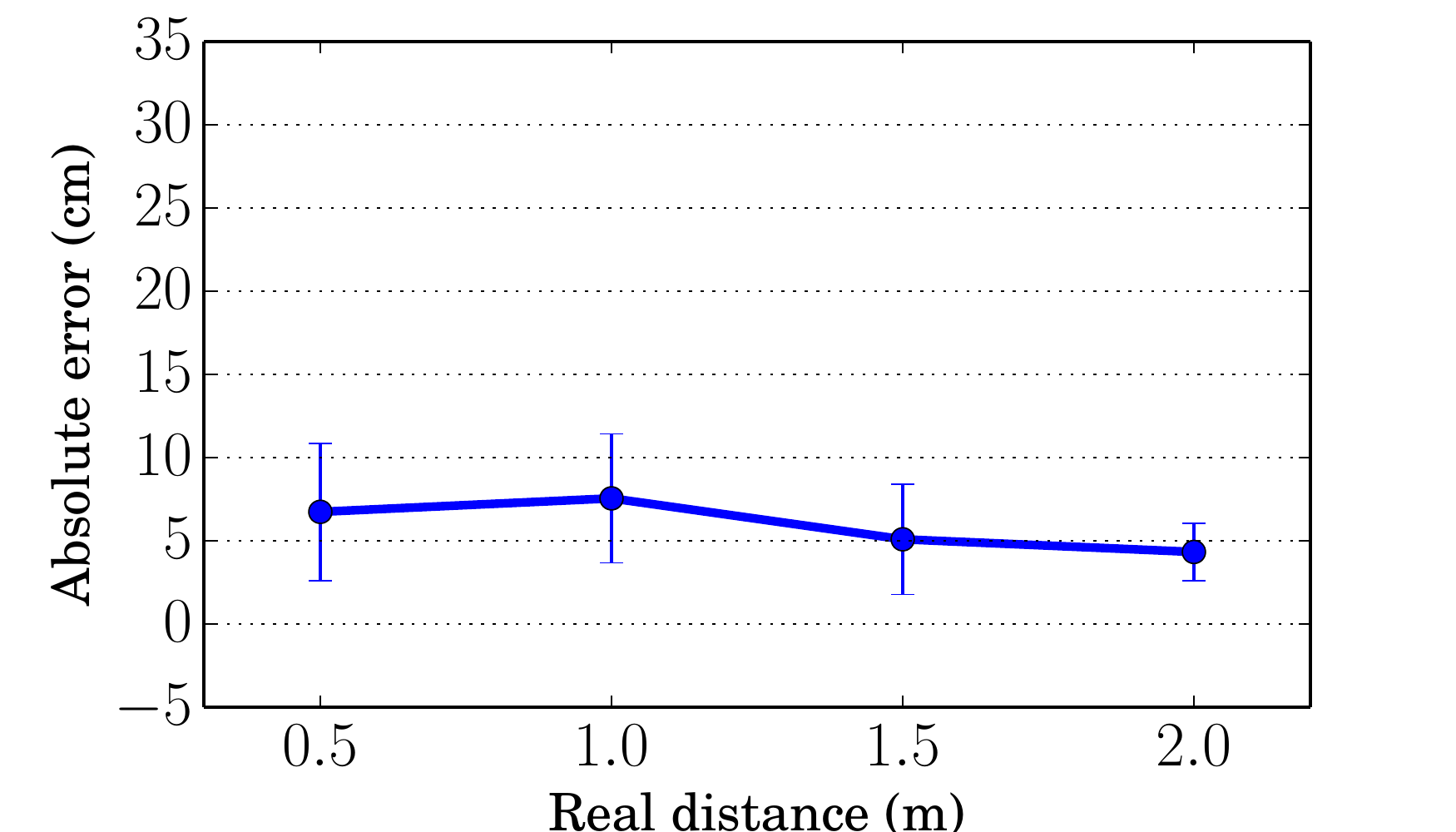}\label{office}}
\subfloat[At home]{\includegraphics[width=0.25 \textwidth]{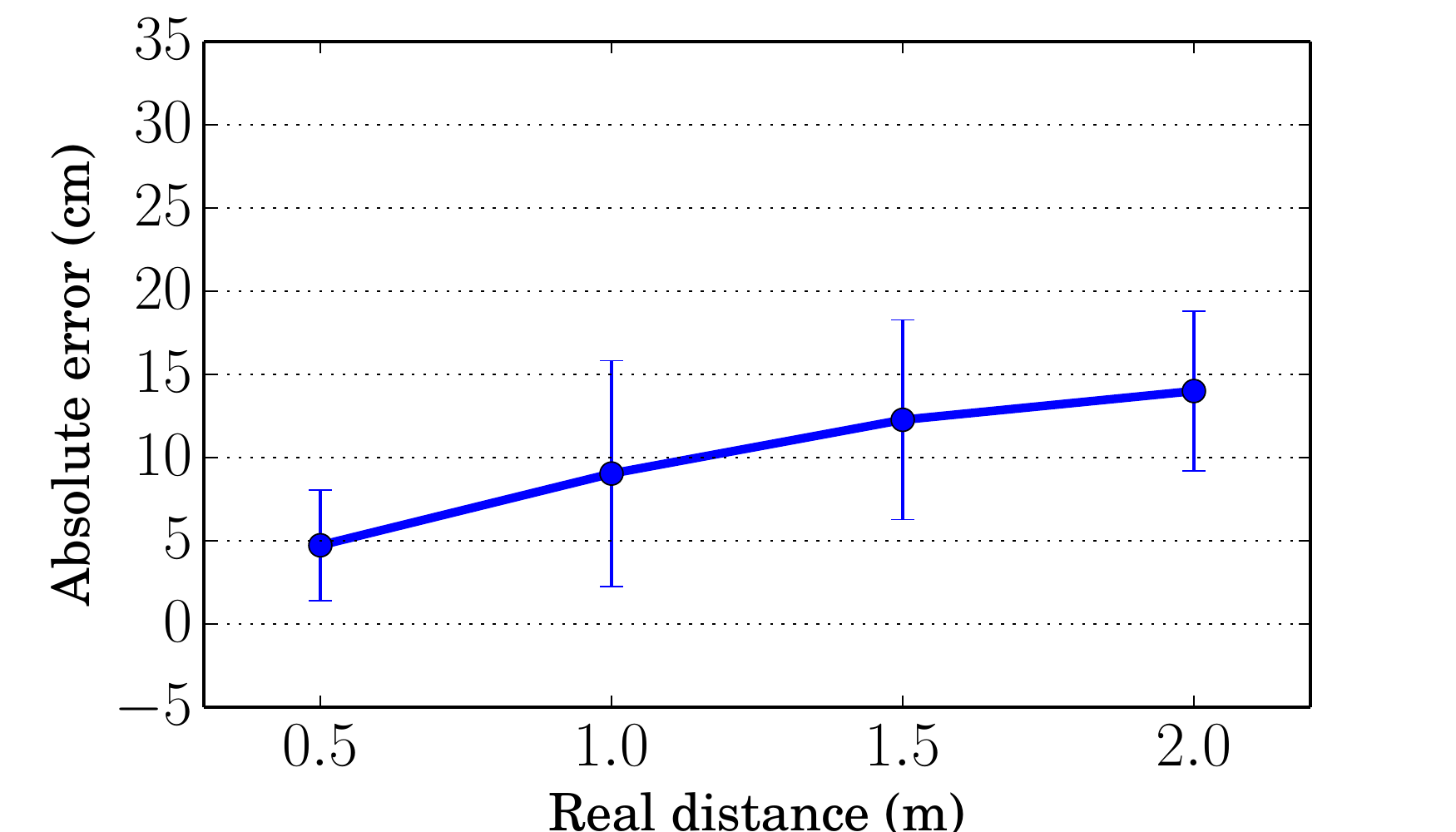}\label{home}}
%\vspace{-2mm}

\subfloat[On the street]{\includegraphics[width=0.25 \textwidth]{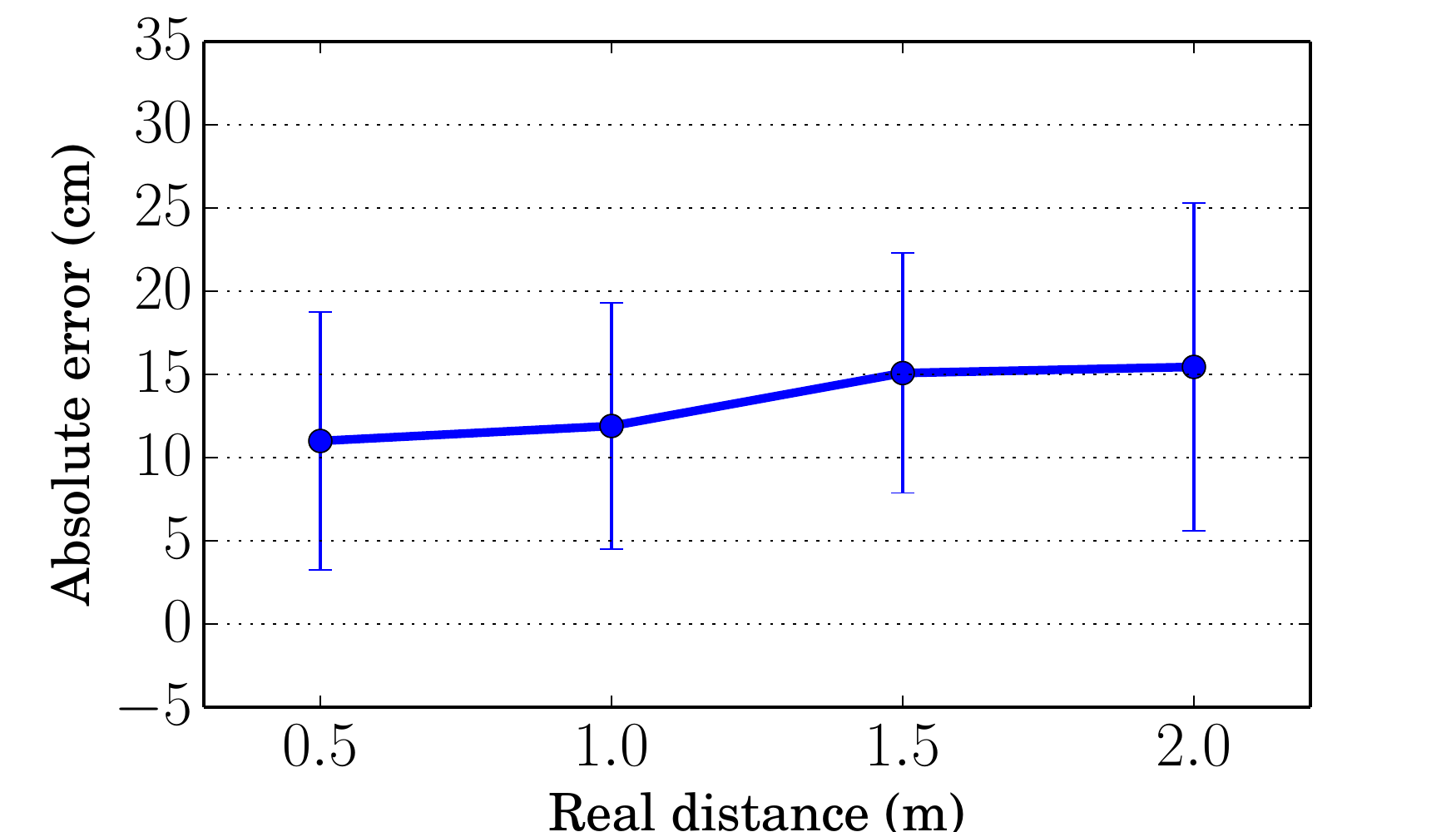}\label{street}}
\subfloat[In a restaurant]{\includegraphics[width=0.25 \textwidth]{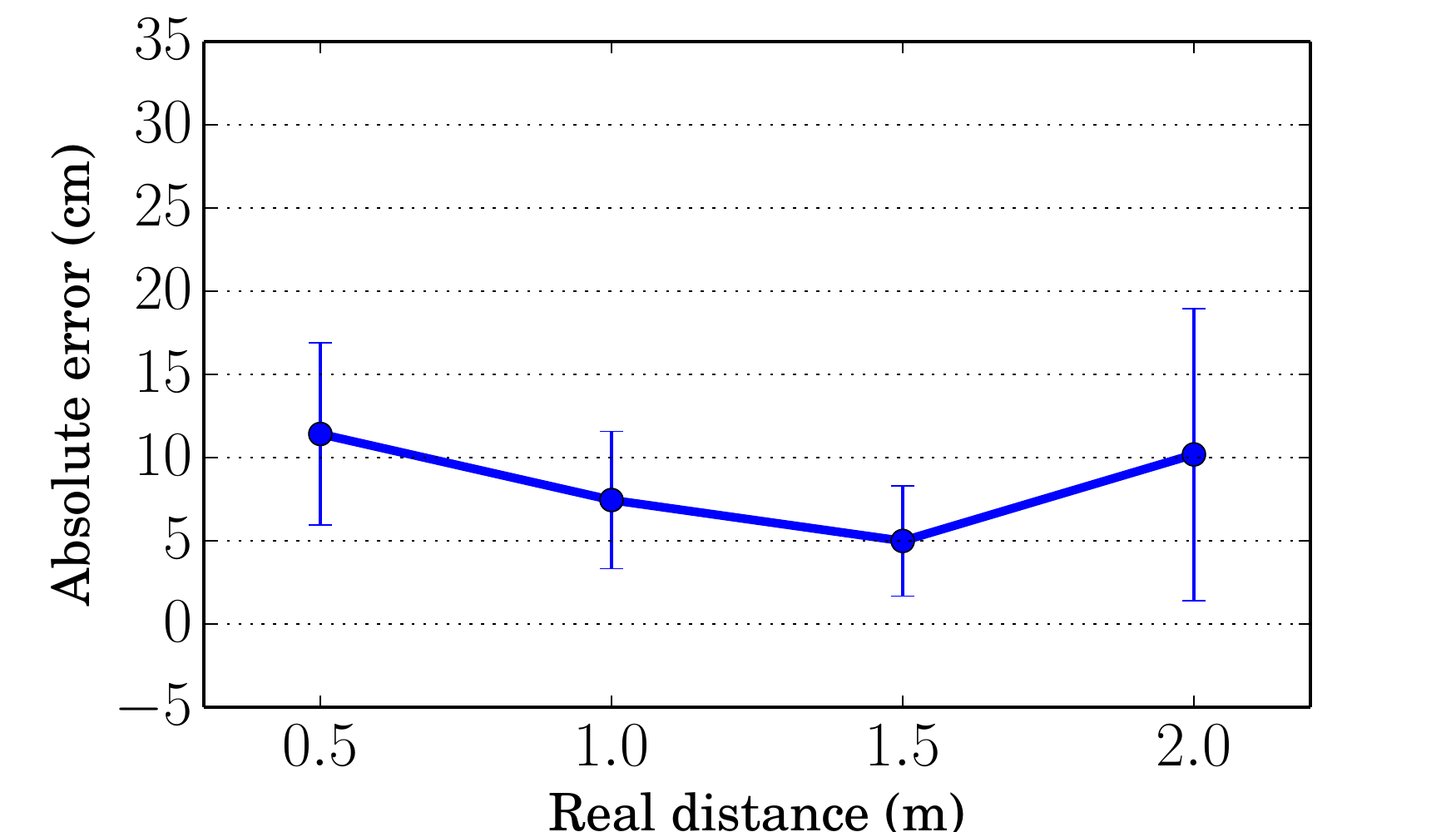}\label{rest}}
\caption{Distance estimation errors in different  environments.}
\label{environment}
\vspace{-4mm}
\end{figure}

\subsection{Accuracy of \SADE\ at Estimating Distance}
\label{comparewithCC}
%We measure the accuracy of our distance estimation protocol \SADE, which is a key component of PIANO.
With the current parameter setting of our prototype, we find that when the real distance between the two devices is larger than around 2.5 meters, ACTION determines that the reference signal is not present in the recorded signal and thus authentication on the authenticating device is denied. Therefore, we measure the accuracy of distance estimation when the real distance is smaller than 2.5 meters. 
Suppose the real distance between the authenticating device and the vouching device is $d$, and the distance estimated by ACTION is $d_{\P\A}$, we define the \emph{absolute error} as $|d-d_{\P\A}|$. We report the absolute errors of distance estimation in different scenarios. For each real distance, we average the absolute errors over 10 trials.

%Figure~\ref{realdistance} illustrates how we measure the real distance between the two smartphones. 
%Each phone has a speaker and a microphone. We denote the two speakers as $S_1$ and $S_2$, and the two microphones as $M_1$ and $M_2$. 
%We treat the real distance as the distance between the speaker $S_1$ and the microphone $M_2$ minus the distance between the speaker $S_1$ and the microphone $M_1$, as it is the distance that ACTION aims to estimate. 

%\begin{figure}[t]
%\center
%{\includegraphics[width=0.45 \textwidth]{3_1_10_Jamming.pdf}}
%\caption{Error bar of distance estimation when an attacker tries to jam our authentication system by playing a acoustic signal consisting of all frequencies in the considered frequency range . We performed the experiments in an office.}
%\label{jam}
%\end{figure}

\subsubsection{Different Environments} 
We evaluate the accuracy of distance estimations in various environments. 

\myparatight{In a shared office, at home, on the street, and in a restaurant} These environments represent places where a user could use PIANO in daily life, and they represent different levels of background noises. For instance, on the street, we have background noise introduced by cars and passersby. In a restaurant, people are chatting and having meals. Figure~\ref{environment} shows the error bars of distance estimation in these environments when the real distance is 0.5, 1, 1.5, and 2 meters. We observe that distance estimation in ACTION is accurate in different environments. Specifically, in a shared office, the average absolute errors are between only 5 centimeters and 7 centimeters. Although the average absolute errors are larger on the street where the background noise is heavier, they vary between 10  and 15 centimeters.

\myparatight{Separated by a wall} %Radio signals (e.g., Wi-Fi, Bluetooth) can go through walls. 
%When the authenticating device and the vouching device are separated by a wall, the authenticating device might be out of the sight of the legitimate user, and an attacker could try to use the authenticating device via performing zero-effort attacks or spoofing attacks. 
%However, if two devices are in two different rooms that are next to each other or two floors that are next to each other,  radio signal based distance estimation/bounding protocols~\cite{kotaru2015spotfi,xiong2015tonetrack, brands1993distance,realizationDistanceBounding,vasisht2016decimeter} would measure their distance to be small, and thus an attacker can use the authenticating device when the proximity-based authentication system  uses a radio signal based distance estimation/bounding protocol.  
%We measure whether our distance estimation protocol still estimates the distance to be small when the two devices are close but are separated by a wall. 
 %We also tried to measure the accuracy when the two smartphones are close but there is a wall between them. 
 We find that, when the two devices are close but are separated by a wall,  one device detects that the reference signal played by the other device is not present, and thus the access to the authenticating device is denied. This is because the reference signals are significantly attenuated by the wall. This means that an attacker cannot access the authenticating device if the user, who carries the vouching device, is separated with the authenticating device by a wall. 
 %When the authenticating device and the vouching device are separated by a wall, the authenticating device might be out of the sight of the user, and an attacker could try to access the authenticating device via performing zero-effort attacks or spoofing attacks. Our results show that, in such cases, the attacker will be blocked. 

%\begin{figure*}[t]
%%\vspace{-2mm}
%\center
%\subfloat[]{\includegraphics[width=0.45 \textwidth]{3_1_10_multi_user.pdf}\label{multiple}}
%\subfloat[]{\includegraphics[width=0.45 \textwidth]{3_1_10_Auto.pdf}\label{comparison}}
%\caption{(a) Error bar of distance estimation when three users are using the authentication system on their devices simultaneously in a shared office. (b) Comparing our frequency-based algorithm with cross-correlation algorithm.}
%%\vspace{-3mm} 
%\end{figure}

% \begin{figure}[t]
%%\vspace{-2mm}
%\center
%\subfloat[]{\includegraphics[width=0.25 \textwidth]{3_1_10_multi_user.pdf}\label{multiple}}
%\subfloat[]{\includegraphics[width=0.25 \textwidth]{3_1_10_Auto.pdf}\label{comparison}}
%
%\caption{(a) Error bar of distance estimation when three users are using the authentication system on their devices simultaneously in a shared office. (b) Comparing our frequency-based algorithm with cross-correlation algorithm.}
%%\label{multiple}
%%\vspace{-3mm} 
%\end{figure}

% \begin{figure}[t]
%%\vspace{-2mm}
%\center
%{\includegraphics[width=0.45 \textwidth]{3_1_10_Auto.pdf}}
%
%\caption{Comparing our frequency-based algorithm with cross-correlation algorithm.}
%%\label{multiple}
%%\vspace{-3mm} 
%\label{comparison}
%\end{figure}

\subsubsection{Multiple Users}
In a public place such as a shared office and a restaurant, multiple users that have adopted PIANO might launch the system on their devices at close times. We measure the accuracy of distance estimation in such scenarios in a shared office. In particular, we assume there are 3 such users. To simulate such scenarios, in each trial of our experiment, we generate 2 pairs of randomized reference signals, and use two other devices to play them when we launch our authentication system on the two smartphones. We performed such simulations at four real distances (i.e., 0.5, 1.0, 1.5, and 2.0 meters), and for each real distance, we repeat for 10 trials. 

First, if two reference signals overlap significantly, ACTION will determine that  the reference signals are not present in the recorded signal. This is because the overlapped reference signal will fail the sanity check at line~\ref{condition} of Algorithm~\ref{power}. However, the probability of such cases is very small. Indeed, in the 40 trials of our experiments, we only observe 3 trials that are such cases. Figure~\ref{multiple} shows the error bar of distance estimation in the remaining trials.  We observe that, compared to Figure~\ref{office} where only one user is using PIANO in a shared office, the average errors are slightly larger. This is because reference signals played by different users could have partial overlaps, which decreases the accuracy slightly.   

%\begin{figure}[t]
%\center
%{\includegraphics[width=0.45 \textwidth]{3_1_10_Auto.pdf}}
%\caption{Comparing our frequency-based algorithm with cross-correlation algorithm.}
%\label{comparison}
%\vspace{-4mm}
%\end{figure}

%\begin{figure}[t]
%%\vspace{-2mm}
%\center
%{\includegraphics[width=0.45 \textwidth]{3_1_10_Auto.pdf}}
%\caption{Comparing our frequency-based algorithm with cross-correlation algorithm.}
%%\vspace{-3mm} 
%\label{comparison}
%\end{figure}

 \begin{figure}[t]
\vspace{-2mm}
\center
\subfloat[]{\includegraphics[width=0.25 \textwidth]{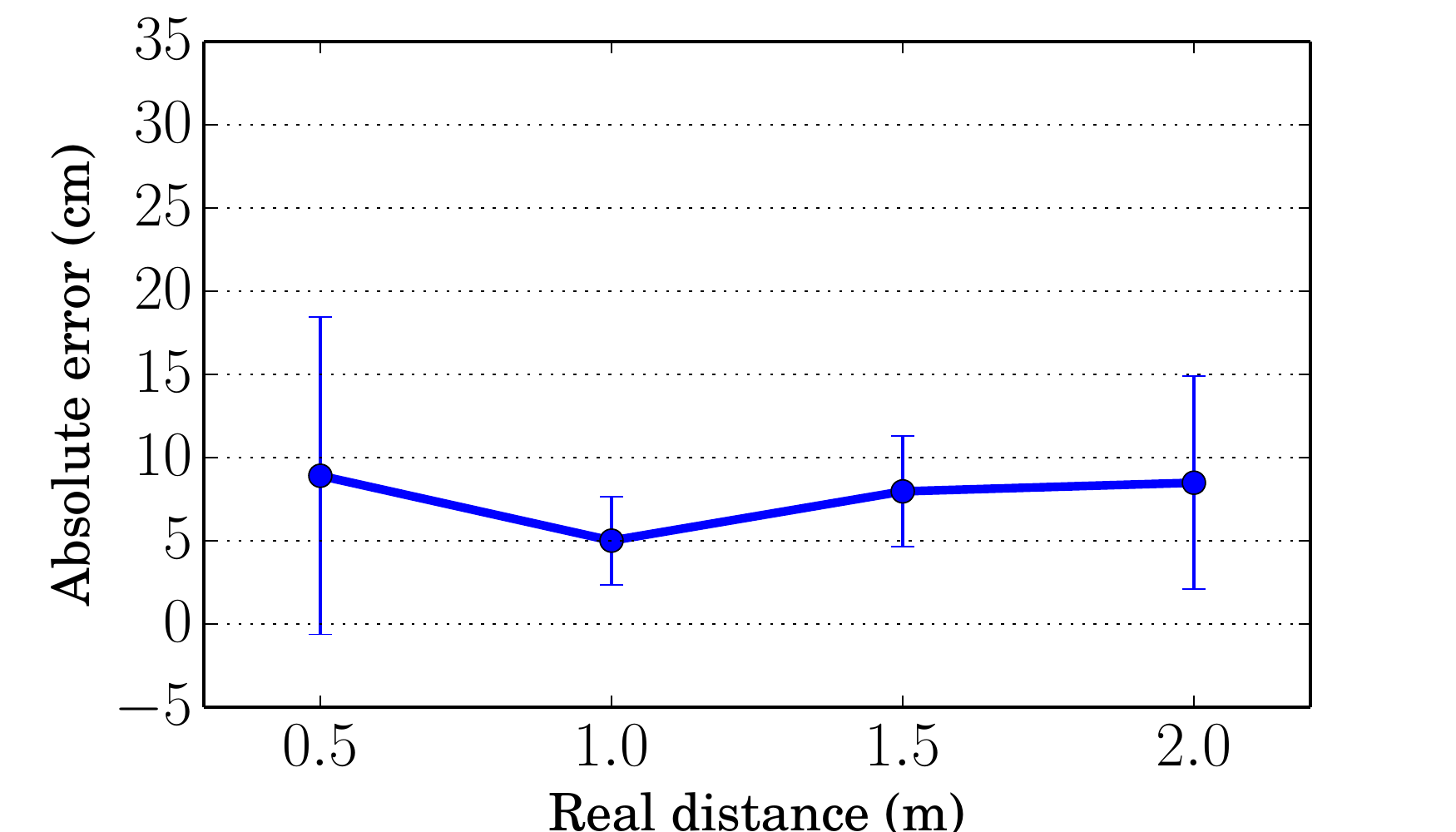}\label{multiple}}
\subfloat[]{\includegraphics[width=0.25 \textwidth]{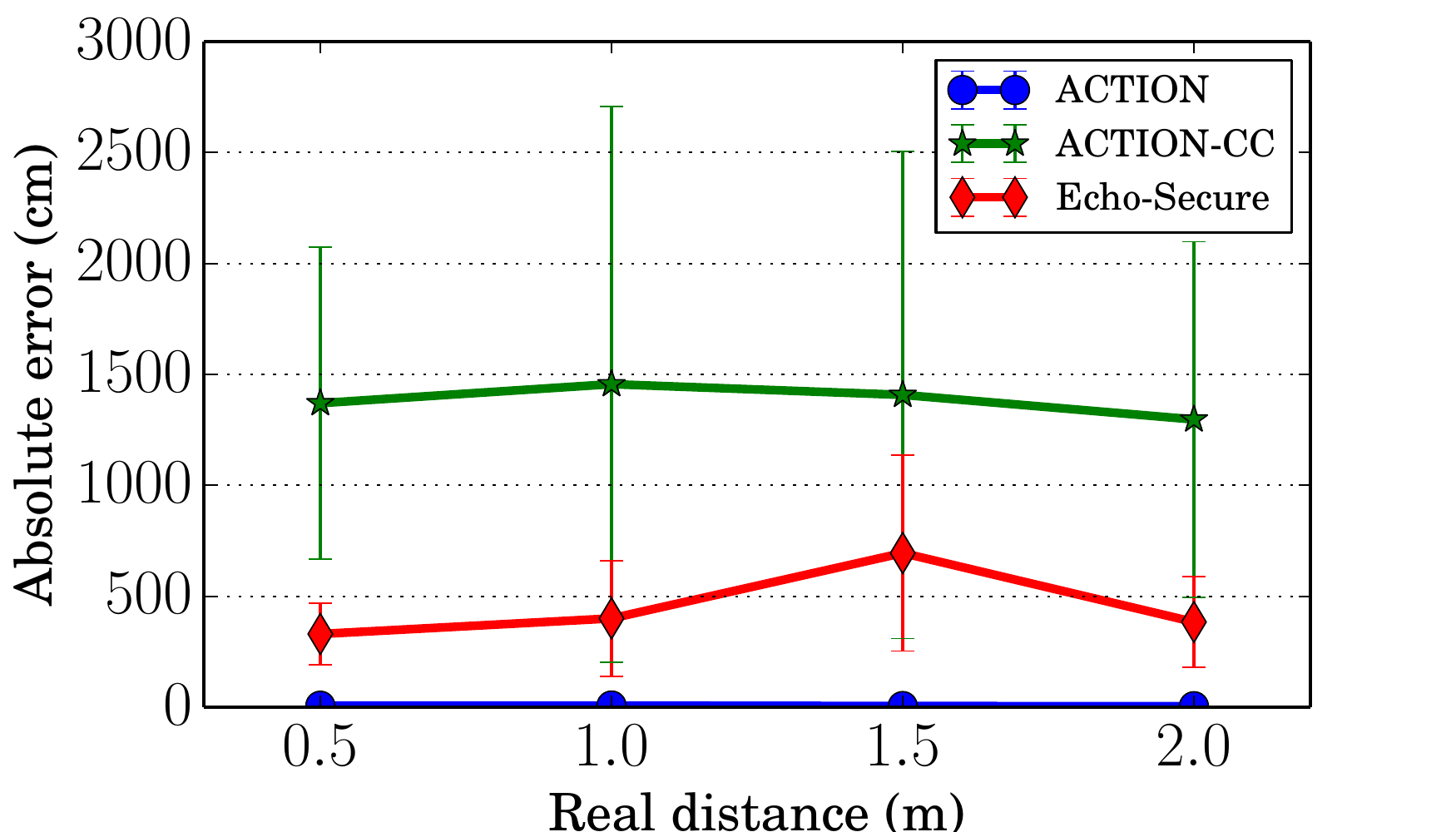}\label{comparison}}
\caption{(a) Error bar of distance estimation when three users are using the authentication system on their devices simultaneously in a shared office. (b) Comparing three secure acoustic signal based distance estimation protocols.}
%\label{multiple}
\vspace{-4mm} 
\label{multiple}
\end{figure}

\subsubsection{Comparison with Previous Methods}
We compare three secure acoustic signal based distance estimation methods: ACTION, ACTION with our frequency-based signal detection algorithm replaced by the cross-correlation algorithm (denoted as ACTION-CC),  
%BeepBeep~\cite{peng2007beepbeep} with the fixed reference signals replaced by our randomized reference signals (denoted as BeepBeep-Secure), 
and Echo~\cite{sastry2003secure} with randomized reference signals and our frequency-based signal detection algorithm (denoted as Echo-Secure). 
In contrast to Echo, 
 Echo-Secure is secure against replay attacks. %Similar to BeepBeep, BeepBeep-Secure also uses the cross-correlation algorithm to detect reference signals. 
Echo was one of the first acoustic signal based distance bounding protocols. When using Echo in our proximity-based authentication, the authenticating device first sends a  reference signal to the vouching device via Bluetooth; the vouching device immediately plays the reference signal after receiving it; the authenticating device records acoustic signals and detects when the reference signal arrived at the authenticating device. Then the distance is the speed of sound multiplies the elapsed time (subtracting the processing delay). We estimated the average processing delay via putting the two devices together (real distance is close to 0) and treating the elapsed time as the processing delay.
The original Echo protocol uses fixed reference signal and does not discuss particular signal detection algorithm. In Echo-Secure, we use randomized reference signal, and our frequency-based algorithm to detect reference signals.

Figure~\ref{comparison} shows the results of the three methods. We performed the experiments in a shared office. We observe that ACTION is orders of magnitude more accurate than ACTION-CC and Echo-Secure. This implies that 1) our frequency-based algorithm is much more accurate  than the cross-correlation algorithm at detecting our randomized reference signals, and 2) processing delays on the devices are unpredictable. 
%Specifically,  average  errors of ACTION-CC are around 14 meters, which is two orders of magnitude higher than the errors achieved by  our frequency-based algorithm. 
 Specifically, ACTION-CC is inaccurate because the reference signals change significantly in the time domain after they are played and recorded, due to 
 frequency smoothing. As a result, cross-correlation algorithm tries to match the original reference signal with the changed reference signal, resulting in high errors. 
 %We note that the original BeepBeep is accurate on commodity devices with a manually designed \emph{fixed} reference signal~\cite{peng2007beepbeep}, but it is vulnerable to replay attacks. 
 Echo-Secure is inaccurate because  processing delay is very unpredictable on the devices. For instance, when the vouching device wants to play the reference signal, there is an unpredictable delay between the API to play acoustic signal is called and the signal is actually played.

\subsection{FRRs and FARs of Authentication}
\label{accuracy}
In the above section, we studied the accuracy of estimating a given distance. In this section, we show  the accuracy--\emph{False Rejection Rate (FRR)} and \emph{False Acceptance Rate (FAR)}-- of authentication decisions made by PIANO.  
We denote by $d_s$ the maximum distance  at which the reference signal played by one device can reach to the other. With our current parameter setting, we have $d_s\approx 2.5$ meters. When the real distance between the two devices is no less than $d_s$, PIANO rejects the access without estimating the distance.  Moreover,   given a real distance $d$ ($0 < d < d_s$) between the two devices, we assume the distance estimated by PIANO follows a Gaussian distribution whose mean is  the real distance $d$  and standard deviation is $\sigma_d$. We note that this assumption does not contradict with our results in the previous section because those distance estimation errors are \emph{absolute errors}. Indeed, using our collected data, we verified that the average estimated distance is very close to the  real distance.  Furthermore, we consider $\sigma_d$ to be constant and we estimate it by averaging the standard deviations at the four points (i.e., 0.5, 1.0, 1.5, and 2.0) obtained in our experiments. Under these settings, we compute FRR by averaging the FRRs at each legitimate distance (i.e., $\leq \tau$) and compute FAR by averaging the FARs at each illegitimate distance ($> \tau$).

Note that we use Bluetooth to pair the two devices. Therefore, 
FAR is 0 when the real distance between the two devices is larger than the communication range of Bluetooth.
In other words, FAR is 0 when the real distance between the two devices is larger than 10 meters, which is roughly the communication range of Bluetooth on many commodity mobile devices~\cite{bluetooth}. 

%the working range of our prototype is around $10$ meters  because we use Bluetooth to pair two devices and the communication range of Bluetooth is around 10 meters on many mobile devices~\cite{bluetooth}. Therefore, FPR is 0 when the real distance between the two devices is larger than 10 meters. 

Table~\ref{table:fnr} and Table~\ref{table:fpr} show the FRRs and FARs (when the two devices are within the communication range of Bluetooth) in different scenarios and for different authentication thresholds.  First, PIANO achieves low FRRs and very low FARs. For instance, in a shared office, FRR is 2.8\% and FAR is 0.3\% when the authentication threshold is 1.0 meter. 
Second, we observe that FRRs decrease quickly while FARs slightly increase as authentication threshold increases. For instance, FRRs decrease by a half in all scenarios when the authentication threshold increases from 0.5 to 1.0 meter. 

\begin{table}[t]\renewcommand{\arraystretch}{1}
\vspace{-2mm}
\centering
\caption{FRRs in different scenarios.}
\begin{tabular}{|c|c|c|c|c|} \hline 
&{ \small 0.5m }&{ \small  1.0m }&{ \small  1.5m }&{ \small 2.0m} \\ \hline
{  \small Office }& { \small 5.6\% }&{ \small  2.8\% }&{ \small  1.9\% }&{  \small 1.4\%} \\ \hline
{ \small Home }& { \small 9.5\% }&{  \small 4.8\% }&{  \small 3.2\% }&{ \small  2.4\%} \\ \hline
{\small  Street }& { \small 12.6\% }&{  \small 6.3\% }&{  \small 4.2\% }&{  \small 3.1\%} \\ \hline
{\small  Restaurant }& { \small 8.5\% }&{ \small  4.2\% }&{  \small 2.8\% }&{\small   2.1\%} \\ \hline
{ \small Multiple users} &{ \small  7.9\% }&{  \small 4.0\% }&{ \small  2.6\% }&{ \small  2.0\%} \\ \hline
\end{tabular}
\label{table:fnr}
\vspace{-4mm}
\end{table}

\subsection{Efficiency}
We measure the efficiency of our prototype in terms of both time and energy consumption. PIANO is fast. In our current implementation, one authentication can be finished within around 3 seconds. We stress that our prototype is just a proof-of-concept. There are various ways to optimize the time efficiency. For example, we can predict when a device will be used, e.g., when  accelerometer and gyroscope data are available, we can detect a device is picked up. Therefore, we can perform authentication before the device is used. Moreover, we use a tool called PowerTutor~\cite{powerTutor} to measure the energy consumption of our prototype. PowerTutor measures the battery energy consumed by an Android app during a period of time. We find that performing 100 times of authentication only consumes 0.6\% of the smartphone battery.

\subsection{Security against Spoofing Attacks}
%We performed these experiments in an office.

%\myparatight{Guessing-based replay attacks} 
We performed 100 trials of guessing-based replay attacks and all-frequency-based spoofing attacks that we discussed in Section~\ref{securityanalysis}. In all of these trials, ACTION detects that the reference signals are not in the recorded signal because of the sanity check at line~\ref{condition} in Algorithm~\ref{power} and line~\ref{lineepsilon} in Algorithm~\ref{frequency}. As a result, all these attack trials failed. 

\section{Conclusion and Future Work}

We propose PIANO, a proximity-based user authentication method for voice-powered IoT devices. PIANO propagates a user's identity from its vouching device to an authenticating device. 
%In PIANO, a user carries a vouching device, which represents the user's identity, and PIANO essentially propagates the identity to the user's IoT devices. PIANO leverages the hardware that voice-powered IoT devices often already have.
The key component of PIANO is a new acoustic signal based protocol that can estimate distance between two devices accurately, efficiently, and securely. 
%We demonstrate that previous acoustic signal based distance estimation protocols are vulnerable to replay attacks, while our protocol is secure against them and other spoofing attacks. 
Via empirical evaluations, we find that our distance estimation protocol is accurate; PIANO achieves low FRRs and FARs at making authentication decisions;  PIANO is fast and has low energy consumption; and PIANO is secure against various spoofing attacks. 
Interesting directions for future work include adapting PIANO to other application scenarios, e.g., web authentication.
% and applying PIANO to more applications such as automatically unlocking doors in hotel.  

\myparatight{Acknowledgement}  We thank Grant Ho and anonymous
reviewers for helpful comments. This material is supported
by the National Science Foundation under Grants No. TWC-1409915, CNS-1238959, CNS-1238962, CNS-1239054, and CNS-1239166.
Any opinions, findings and conclusions or recommendations
expressed in this material are those of the author(s) and do
not necessarily reflect the views of the funding agencies.

\begin{table}[!t]\renewcommand{\arraystretch}{1}
\centering
\vspace{-2mm}
\caption{FARs  in different scenarios.}
\begin{tabular}{|c|c|c|c|c|} \hline 
&{ \small 0.5m }&{ \small  1.0m }&{ \small  1.5m }&{\small  2.0m} \\ \hline
{ \small Office }&{ \small  0.3\% }&{ \small   0.3\% }&{ \small   0.3\% }&{  \small  0.4\%} \\ \hline
{ \small Home }&{ \small   0.5\% }&{ \small   0.5\% }&{ \small   0.6\% }&{ \small   0.6\%} \\ \hline
{ \small Street }&{ \small   0.7\% }&{  \small  0.7\% }&{ \small   0.7\% }&{ \small   0.8\% }\\ \hline
{ \small Restaurant }&{\small    0.4\% }&{  \small  0.5\% }&{\small    0.4\% }&{  \small  0.4\%} \\ \hline
{\small  Multiple users }&{ \small   0.4\% }&{  \small  0.4\% }&{  \small  0.5\% }&{  \small  0.5\%} \\ \hline
\end{tabular}
\label{table:fpr}
\vspace{-4mm}
\end{table}

%%\balance
%{
%%\vspace{6mm}
%\small
%\bibliographystyle{unsrt}
%\bibliography{refs}
%}
%%\input{appendix}

\end{document}